\RequirePackage{lineno} 
\documentclass{nature}
\usepackage{xcolor}
\usepackage{upgreek}
\usepackage{caption}
\usepackage{multirow}
\usepackage{acronym}
\usepackage{array}
\usepackage{amsmath}
\usepackage{bm}
\usepackage{braket}
\usepackage{amsmath, amssymb, amsfonts}

\usepackage{calc}

\title{Evolution of flat band and role of lattice relaxations in twisted bilayer graphene}

\author{Qian Li$^{1,\dagger}$, Hongyun Zhang$^{1,\dagger}$, Yijie Wang$^{2}$, Wanying Chen$^1$, Changhua Bao$^1$, Qinxin Liu$^1$, Tianyun Lin$^1$,  Shuai Zhang$^{3}$, Haoxiong Zhang$^1$, Kenji Watanabe$^4$, Takashi Taniguchi$^5$, Jose Avila$^6$, Pavel Dudin$^6$, Qunyang Li$^{3}$, Pu Yu$^{1,7}$, Wenhui Duan$^{1,7,8}$, Zhida Song$^{2}$ \& Shuyun Zhou$^{1,7,*}$}

\usepackage{graphicx}
\makeatletter
\let\saved@includegraphics\includegraphics
\AtBeginDocument{\let\includegraphics\saved@includegraphics}

\makeatother

\begin{document}

	\maketitle
	\begin{affiliations}
		
		\item State Key Laboratory of Low-Dimensional Quantum Physics and Department of Physics, Tsinghua University, Beijing 100084, People’s Republic of China
		
		\item International Center for Quantum Materials, School of Physics, Peking University, Beijing 100871, People’s Republic of China
		
		\item AML, CNMM, Department of Engineering Mechanics, Tsinghua University, Beijing 100084, People’s Republic of China
		
		\item Research Center for Functional Materials, National Institute for Materials Science, 1-1 Namiki, Tsukuba 305-0044, Japan
		
		\item International Center for Materials Nanoarchitectonics, 
		National Institute for Materials Science, 1-1 Namiki, Tsukuba 305-0044, Japan
		\item Synchrotron SOLEIL, L’Orme des Merisiers, Saint Aubin-BP 48, 91192 Gif sur Yvette Cedex, France
		\item Frontier Science Center for Quantum Information, Beijing 100084, People’s Republic of China
		\item Institute for Advanced Study, Tsinghua University, Beijing 100084, People’s Republic of China
		
		$\dagger$ These authors contributed equally to this work.
		
		* Correspondence should be sent to syzhou@mail.tsinghua.edu.cn.
	\end{affiliations}
	
	\newpage
	
	\begin{abstract}
		Magic-angle twisted bilayer graphene (MATBG) exhibits correlated phenomena such as superconductivity and Mott insulating state related to the weakly dispersing flat band near the Fermi energy. Beyond its moiré period, such flat band is expected to be sensitive to lattice relaxations. Thus, clarifying the evolution of the electronic structure with twist angle is critical for understanding the physics of MATBG. Here, we combine nanospot angle-resolved photoemission spectroscopy and atomic force microscopy to resolve the fine electronic structure of the flat band and remote bands, and their evolution with twist angles from 1.07$^\circ$ to 2.60$^\circ$. Near the magic angle, dispersion is characterized by a flat band near the Fermi energy with a strongly reduced bandwidth. Moreover, near 1.07$^\circ$, we observe a spectral weight transfer between remote bands at higher binding energy and extract the modulated interlayer spacing near the magic angle. Our work provides direct spectroscopic information on flat band physics and highlights the role of lattice relaxations.
	\end{abstract}
	
	\newpage
	
	\renewcommand{\thefigure}{\textbf{Fig. \arabic{figure} $\bm{|}$}}
	\setcounter{figure}{0}

	The emergence of flat band near the Fermi energy $E_F$, where electrons are compressed into a small energy range thereby enhancing the electron-electron interaction\cite{bistritzer2011moire,dos2007graphene,dos2012continuum,de2012numerical}, is fundamental to the correlated phenomena reported in magic-angle twisted bilayer graphene (MATBG)\cite{cao2018unconventional,cao2018correlated}.    
	The flat band depends sensitively not only on the  moir\'e superlattice, but also on the lattice relaxations\cite{KoshinoPRB2017,FabrizioPRB2018,VishwanathPRL2019,GiovanniPRB2019,WaletPRB2019,KaxirasPRR2019,LucignanoPRR2020,Kaxiras2020,RubioRev2021} as a result to minimize the total energy, because different local stackings such as AA, AB (schematically illustrated in the inset of Fig.~1a) have different energies\cite{KaxirasPRR2019}. It was theoretically predicted that lattice relaxations could strongly affect the electronic structure, for example, they are indispensable for opening up an energy gap, separating the flat band near $E_F$ from the remote bands at higher binding energy\cite{GiovanniPRB2019,LucignanoPRR2020}. 
	
	Experimentally, lattice relaxations in MATBG are investigated by local probes such as scanning tunneling microscopy (STM)\cite{Pasupathy2019maximized}, atomic force microscopy (AFM)\cite{PasupathPFMNatNano2020,LiQYSciAdv2020,LiQY2011atomic} and transmission electron microscopy (TEM)\cite{KimNatMater2019,BediakoStrainNatMater2021}, and the flat band is revealed by spectroscopic probes such as scanning tunneling spectroscopy (STS)\cite{Pasupathy2019maximized,Yazdani2019spectroscopic,Eva2019ChargeOrder, NadjPerge2019electronic} and nanospot angle-resolved photoemission spectroscopy (NanoARPES) measurements\cite{BaumbergerARPES2021,WangFARPES2021}. However, so far, the direct correlation of lattice relaxations and the actual electronic structure is still elusive. Experimental evolution of the energy- and momentum-resolved electronic structure with precisely controlled twist angle is highly demanded and critical for revealing the interlayer coupling and the role of lattice relaxations, however, so far this has remained missing.
	Here by combining two complementary spectroscopic and microscopic experimental probes - NanoARPES\cite{JozwiakNatPhys2018,Kondo2019weak,Xu2019visualizing,bao2017stacking} which allows to probe the electronic structure\cite{DamascelliRMP2003,zhang2022angle} and AFM\cite{PasupathPFMNatNano2020,LiQYSciAdv2020,LiQY2011atomic} which allows to resolve the crystalline structure of the moir\'e superlattice (Fig.~1a), we report the evolution of the flat band and remote bands with twist angle from 1.07$^\circ$ to 2.60$^\circ$. Combining the advantages of high-quality, momentum-resolved electronic structure, spatially-resolved moir\'e superlattice, together with theoretical calculations, we provide important information on the fundamental physics of twisted bilayer graphene near the magic angle, in particular regarding the interlayer coupling and the critical role of lattice relaxations. 
 
\section*{Experimental overview from NanoARPES and AFM measurements}

	High-quality twisted bilayer graphene (tBLG) samples with desired twist angle were prepared by a ``tear-and-stack'' method using dry-clean transfer\cite{wang2013one} (see Methods and the Extended Data Fig.~1).
	The moir\'e period and twist angle were extracted from AFM measurements using either the lateral force (L-AFM)\cite{LiQY2011atomic} or conductive current (C-AFM)\cite{LiQYSciAdv2020} before or after NanoARPES measurements, during which special care was taken to ensure that AFM and NanoARPES measurements were performed on the same spot (see Extended Data Fig.~2), so that the momentum-resolved electronic structure can be directly correlated with the moir\'e superlattice or twist angle. 
	
	The NanoARPES dispersion image in Fig.~1a shows an isolated flat band (pointed by the red arrow), with a gap clearly separating the flat band from remote bands at higher binding energy (indicated by the blue arrow). Such remote bands have been largely overlooked, while recently it has been suggested that their momentum distribution can be used to extract the interlayer tunneling parameters\cite{MacDonald2021PRB}. 
	Figure 1b-f shows an overview of three-dimensional NanoARPES intensity maps measured on samples at five different twist angles, which are determined to be 1.07$^\circ$ to 2.60$^\circ$ by AFM measurements (Fig.~1g-k, see Extended Data Fig.~3 for more information).
	Interestingly, a spectral weight transfer is observed from a systematic evolution of the remote bands, namely, there is a switching from a stronger intensity for the inner pocket (labeled as $p_1$, pointed by the red arrow in Fig.~1b) at twist angle of 1.07$^\circ$, to a stronger intensity for the outer pocket (labeled as $p_2$, pointed by the blue arrow) at twist angles of 1.31$^\circ$ (Fig.~1c) and larger (Fig.~1d-f). The evolution of the flat band with twist angle and the spectral weight transfer carries important information about the interlayer coupling and lattice relaxations, which are the main focus of this work.

\section*{Flat band and AA- plus AB- like remote bands at magic angle}

	Figure 2 shows the flat band measured on tBLG at twist angle of 1.07$^\circ$. Figure 2a shows a schematic for the moir\'e superlattice Brillouin zone (mSBZ), where the Dirac points from top and bottom graphene layers are labeled by $K_t$ and $K_b$ respectively. Figure 2b-e shows dispersion images measured by cutting  through the flat band along different directions. The weakly dispersive flat band is observed for all these cuts, as pointed by red arrows. The dispersion of the flat band and remote bands can be extracted by following peaks in the energy distribution curves (EDCs) shown in Fig.~2f-i, where peaks from the flat band are indicated by red tick marks, while those from the remote bands are indicated by black and blue tick marks. 
	From the extracted dispersions of the flat band in Fig.~2j, the bandwidth of the flat band is extracted to be 11 $\pm$ 10 meV, which is comparable to theoretical predictions\cite{bistritzer2011moire} as well as results obtained from the energy width of the $dI/dV$ peak in STM measurements\cite{Pasupathy2019maximized}, while here the high-quality NanoARPES data allow us to directly visualize the dispersions of the flat band and remote bands along different directions with momentum-resolved information.  
	Figure 2k shows the extracted dispersions for cut e, which mainly consist of the flat band near $E_F$ (red curve), remote bands (black and gray curves) resembling those from AB stacking, and an additional band between -0.35 eV and $E_F$ (blue curve) which is a characteristic dispersion feature from AA stacking. The dispersions of the remote bands show overall similarity to AA-like plus AB-like dispersions (see simulated dispersions in Fig.~2l). The above analysis shows that the electronic structure near the magic angle is characterized by the flat band with a narrow bandwidth of approximately 11 meV, and AA-like plus AB-like graphene remote bands, exhibiting notable difference from dispersions at larger twist angles (see Supplementary information Fig.~S1 for the dispersion analysis of 2.22$^\circ$ tBLG).

	In order to further reveal how the electronic structure near the magic angle evolves with twist angle, we show in Fig.~3a-e dispersion images measured through $K_t$ and $K_b$ (indicated by the black line in the inset of Fig.~3a) at different twist angles. The flat band shows up as weak intensity near $E_F$ at twist angle of 1.07$^\circ$ (red dotted curve), and it becomes more pronounced with increasing twist angle (Fig.~3b-e). The extracted dispersions of the flat band from EDC analysis are over-plotted as red curves (see Extended Data Fig.~4 for the extraction of the dispersion from EDC analysis), which reveal an evolution from weakly dispersive flat band at 1.07$^\circ$ to a clear M-shaped dispersion at 2.60$^\circ$, indicating a strong increase of the bandwidth with increasing twist angle. Figure 3f-j shows calculated dispersions (see more information about the theoretical model in Methods), where gaps are observed at two momenta indicated by dashed lines. A comparison of the extracted bandwidth from experimental results and calculations in Fig.~3k shows a good agreement, with an increase of the experimental bandwidth from 11 $\pm$ 10 meV at 1.07$^\circ$ to 215 $\pm$ 30 meV at 2.60$^\circ$. 
	The corresponding Coulomb energy calculated by $U_{eff} = e^2 / (4\pi\epsilon_0\epsilon_r\lambda_m)$  for different twist angles is also plotted in Fig. 3k (black dashed line)\cite{cao2018correlated}, where $\epsilon_0$ and $\epsilon_r$ refer to the dielectric constants of the vacuum and the substrate respectively, and $\lambda_m$ is the corresponding moir\'e period. Near the magic angle, the bandwidth is smaller than the Coulomb energy, which is in line with the maximized electron interaction from STM measurements\cite{Pasupathy2019maximized}.

	In addition to the flat band, the remote bands also carry important information about the interlayer coupling parameters, such as the interlayer tunneling amplitude for AB stacking $\omega_{AB}$, and the ratio of the tunneling strength between AA and AB stackings, $\alpha = \omega_{AA}/\omega_{AB}$.  A schematic illustration of the hybridization between the remote bands is shown in Figure~3l, where the interference between the two Dirac cones (gray circles) from the top and bottom graphene layers leads to a splitting of the remote bands into $p_1$ and $p_2$. The separation of the remote bands, both in momentum space and in energy $\Delta E$, can therefore be used to extract $\omega_{AB}$ and $\alpha$ (see more information in the Methods).  A comparison of the experimental and calculated intensity maps at twist angle of 2.22$^\circ$ (Extended Data Fig.~5) suggests that the parameters of $\omega_{AB} $ = 120 meV and $\alpha$ = 0.8 give a momentum separation in agreement with the experimental result. 
	Figure~3m-q shows dispersion images measured along the black line in Fig.~3l, from which the extracted $\Delta E$ is plotted in Fig.~3r (see Supplementary information Fig.~S2 for more detailed analysis).  To extract the tunneling parameters, we have performed theoretical calculations for different twist angles using the BM model (see Methods for more details) with different tunneling parameters, and compared the calculated $\Delta$E with experimental results.  Here we note that  $\Delta E$ depends on the choice of $k_y$ (Extended Data Fig.~6), and therefore $k_y$ needs to be fixed when comparing experimental results and calculated results, or when comparing $\Delta E$ at different twist angles. The comparison in Fig.~3r shows that $\omega_{AB} = 120 \pm 10 $ meV and $\alpha =\omega_{AB} / \omega_{AB} = 0.8\pm0.1$ (diamond symbols in Fig.~3r) give the best agreement with the experimental results. In addition to the modulation in the splitting of remote bands with twist angle, another important observation is that the AA-like remote band observed at twist angle of 1.07$^\circ$ (pointed by gray arrow in Fig.~3m) becomes almost absent at larger twist angles, which is accompanied by a spectral weight transfer between $p_1$ and $p_2$. Below we further analyze the momentum distribution of the spectral weight transfer and discuss the underlying mechanism.

\section*{Spectral weight transfer between different remote bands}

	The spectral weight transfer between the remote bands is observed more clearly from the ARPES intensity maps shown in Fig.~4. 
	The Fermi surface maps in Fig.~4a1-e1 show two point-like Fermi surfaces together with their weak moir\'e replicas, whose separation increases with twist angle, in agreement with the smaller moir\'e period observed in AFM measurements. 
	Moving down in energy, the overall band topology shows a transition from nearly circular-shaped surface topology at 1.07$^\circ$ (Fig.~4a4) to more oval-shaped surface topology at 2.60$^\circ$ (Fig.~4e4).  When increasing the twist angle, a spectral weight transfer from stronger intensity for pocket $p_1$ to stronger intensity for pocket $p_2$ is observed in the momentum-resolved surface topology (Fig.~4a3-e4) and momentum distribution curves (MDCs) in Fig.~4f. We note that the intensity ratio between $p_1$ and $p_2$ is not affected by the measurement positions in the same sample or different samples with the same twist angle (see Extended Data Fig.~7), suggesting that it is an intrinsic feature that depends on the twist angle.

\section*{Evolution of the flat band and remote bands with twist angle}

	To reveal the possible mechanism of the spectral weight transfer, we first compare the calculated results using the Bistritzer-MacDonald model\cite{bistritzer2011moire} with $\omega_{AB}$  and $\alpha$ = $\omega_{AA}/\omega_{AB}$ extracted above. However, while the calculated band structures in Fig.~4g-k can reproduce the overall shapes and positions of the pockets, the lower pocket $p_2$ always has a stronger intensity for all the twist angles calculated, which is in contrast to the experimental results. We note that including the correlation effect\cite{lian2021twisted} only affects the flat band near $E_F$ (see Supplementary information Fig.~S3 for more details) without switching the intensity of the remote bands, suggesting that this cannot be explained by correlation effect. Instead, interestingly, when using different values of interlayer spacing $c$, the calculated surface maps in Fig.~4l-p show spectral weight transfer similar to the experimental results. 

\section*{Critical role of lattice relaxations}
	
	In order to obtain an intuitive picture of the underlying physics of the spectral weight transfer and its connection with the interlayer spacing, we first explain how the ARPES data can reveal the interlayer coupling and lattice relaxation. The ARPES intensity $I(p,E)$ is a quantum coherent interference of the contributions from both sublattices $\alpha$ and layers $l$, which can be written as equation(1) \cite{MacDonald2021PRB}

	\begin{equation}
		\begin{aligned} 
			I(p,E) \propto \sum_{\bm{k},\lambda}\lvert \sum_{l} B_l \sum_{\alpha } A_ \alpha \sum_{G}\delta_{p_\Vert, k+G }\Psi  _{G,l,\alpha;\lambda}(k)\rvert^2 \delta(E-\epsilon_{\bm{k},\lambda})
		\end{aligned} 
	\end{equation}
	
	where $\Psi_{G,l,\alpha;\lambda}(k)$ and  $\epsilon_{\bm{k},\lambda}$ denote the wave-functions and energies of the moiré bands, $A_ \alpha =e^{\pm i\varphi }$ for $\alpha =A,B$ is determined by the in-plane polarization angle $\varphi $ which tunes the interference between these two sublattices, while $B_l =e^{ik_\bot \cdot lc }$ for $l=0,1$ is determined by the out-of-plane photon momentum $ik_\bot $ combined with the average interlayer spacing $c$. In particular, here $B_l$ tunes the interference phase between the two layers, which is the dominant factor leading to the spectral weight transfer as shown in Fig.~4. 
	
	The modulation of the ARPES intensity by the interlayer spacing is testified by the calculated dispersions in Fig.~5a-c for twist angle of 1.07$^\circ$, where by changing $c$ from 3.35 \AA~to 3.60~\AA, a switching from stronger $p_1$ to stronger $p_2$ is observed.  A quantitative analysis of the intensity ratio between $p_1$ and $p_2$ shows that the best agreement with the experimental results is obtained using $c$ = 3.40 \AA~(see Extended Data Fig.~8 for more detailed information). Similarly, the interlayer spacing for other twist angles can be extracted (Extended Data Fig.~9).

	To provide more insights into the evolution of the interlayer spacing and different electronic states with twist angle, we show in Figure 5d-f calculated intensity maps at -0.6 eV for three representative twist angles, 1.07$^\circ$, 1.31$^\circ$ and 2.22$^\circ$ respectively. 
	The corresponding projected local density of states (LDOS) for the flat band near $E_F$, pockets with AA-like and AB-like dispersions ($p_{AA}$ and $p_{AB}$, as indicated by colored dotted ovals in Fig.~5b) are shown in Fig.~5g-o. 
	The comparison shows that near the magic angle, electrons contributing to the flat band become more localized in the AA-stacking region (red dot in Fig.~5g), while in contrast, electrons contributing to the remote bands, $p_{AA}$  and $p_{AB}$, are overall much more delocalized (Fig.~5j, m). In addition, the LDOS for $p_{AA}$ increases when approaching the magic angle (Fig.~5j), which is in agreement with the experimental observation of stronger AA-like dispersion in Fig.~3m. Such localized flat band and delocalized remote bands make it possible to describe the physics in the heavy-fermion  model proposed for MATBG\cite{BernevigPRLheavyfermion2022,DaiXPRB2022}. 
	
	The extracted interlayer spacing $c$ in Fig.~5p can be viewed as a weighted average interlayer spacing between different stacking regions, and the decrease of $c$ near the magic angle suggests that there are significant lattice relaxations as schematically illustrated in Fig.~5q. Such lattice relaxations result in smaller AA stacking areas which become more isolated near the magic angle, which is supported by C-AFM measurements\cite{LiQYSciAdv2020} and the calculated LDOS in Fig.~5g, meanwhile, the increase of AB stacking areas makes $c$ approach the interlayer spacing for AB stacking, 3.35 \AA. The above analysis is also supported by the vertical local conductivity using C-AFM measurements (see Extended Data Fig.~10), which provide complementary evidence that the region between two adjacent AA-stacking regions does not fully relax to AB-stacking until at small twist angles near the magic angle. Such lattice relaxations or buckling can be effectively viewed as applying a pseudo-magnetic field or confined quantum dots\cite{GuineaPRB2008,AndreiNature2020}, which is in line with the pseudo-magnetic field interpretation of MATBG\cite{DaiXPLLPRB2019,ZhangZYNC2020}.  Finally, we note that experimental evolution of the electronic structure with controlled twist angle has been in demand\cite{MacDonald2021PRB,RubioRev2021}, and our work shows that it can indeed provide rich information, in particular such spectroscopic evolution highlights the role of lattice relaxations in the electronic structure of MATBG.

	\begin{addendum}

		\item[Acknowledgement] This work is supported by the National Key R$\&$D Program of China (No.~2021YFA1400100, 2020YFA0308800), the National Natural Science Foundation of China (No.~12234011, 52025024, 92250305,12327805,12304226), the Basic Science Center Program of NSFC (No.~52388201). 
		H. Z. acknowledges support from the Shuimu Tsinghua Scholar project and the Project funded by China Postdoctoral Science Foundation (Grant No.~2022M721887). K.W. and T.T. acknowledge support from the Elemental Strategy Initiative conducted by the MEXT, Japan, Grant Number JPMXP0112101001, JSPS KAKENHI Grant Number JP20H00354 and the CREST(JPMJCR15F3), JST. We acknowledge SOLEIL for the provision of synchrotron radiation facilities.
		
		\item[Author Contributions] S.Z. designed the research project. Qian L. and H.Z. prepared the samples. H.Z., Qian L., W.C., J.A., P.D. and S.Z. performed the NanoARPES measurements and analyzed the ARPES data. Qian L. performed the AFM measurements, with assistance from Shuai Z., Qunyang. L. and P. Yu. K.W. and T.T. grew the BN crystals.  Y.W. and Z.S. performed the numerical calculations. Q.L., H.Z. P.Y. and S.Z. wrote the manuscript, and all authors commented on the manuscript.
		
		\item[Competing Interests] The authors declare that they have no competing financial interests.
		
		\item[Supplementary information] is available for this paper. 
		
		\item[Correspondence and requests for materials] should be addressed to Shuyun Zhou (email: syzhou@mail.tsinghua.edu.cn).
		
	\end{addendum}

	\begin{figure*}[htbp]
		\centering
		\includegraphics[width=16.8 cm]{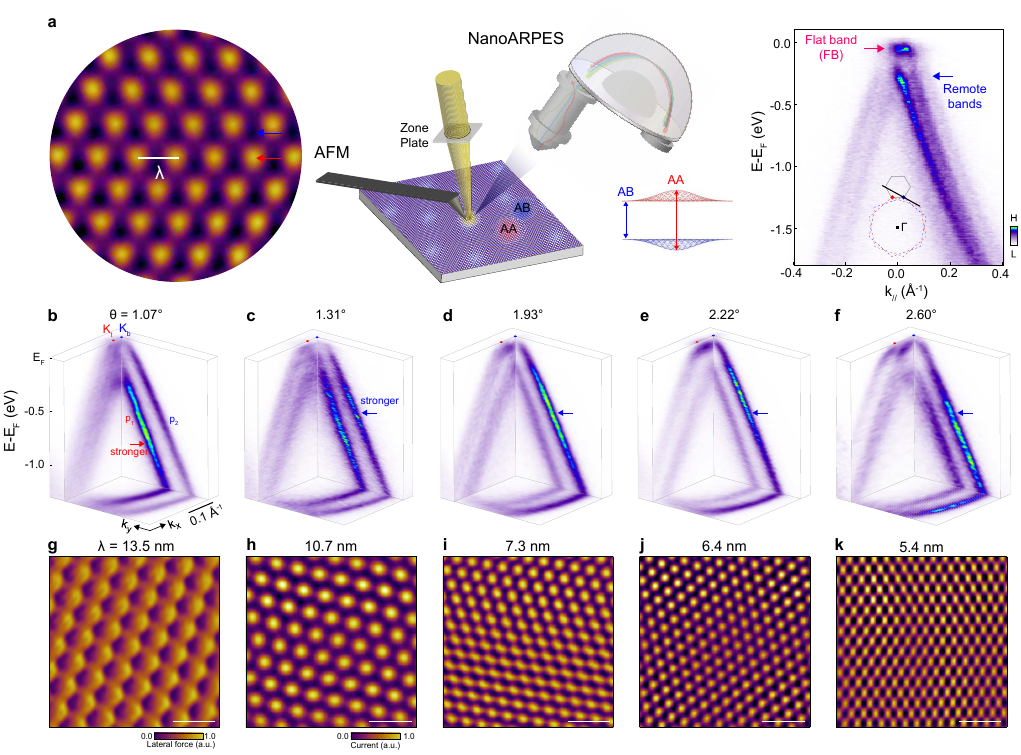}
		\caption{\textbf{Resolving the electronic structure and moir\'e superlattice by combining NanoARPES and AFM measurements.} \textbf{a}, Schematic illustrations of  AFM and NanoARPES  measurements. The plot on the left shows the moir\'e superlattice from AFM measurements, and the plot on the right shows the flat band (pointed by red arrow) and remote bands (pointed by blue arrow) revealed by NanoARPES measurements. The inset on the right shows the moir\'e superlattice Brillouin zone (mSBZ, gray hexagon) and graphene Brillouin zone (BZ) from the top (red) and bottom (blue) graphene layers. \textbf{b-f}, Three-dimensional band structures for twist angle ranging from 1.07$^\circ$ to 2.60$^\circ$. Red and blue dots correspond to the Dirac points from the top ($K_t$) and bottom ($K_b$) graphene layers, respectively. \textbf{g-k}, Moiré superlattices revealed by AFM measurements, using the L-AFM (\textbf{g,k}) or C-AFM (\textbf{h-j}) modes. The scale bars in \textbf{g-k} are all 20 nm.}
	\end{figure*}
	
	\begin{figure*}[htbp]
		\centering
		\includegraphics[width=16.8 cm]{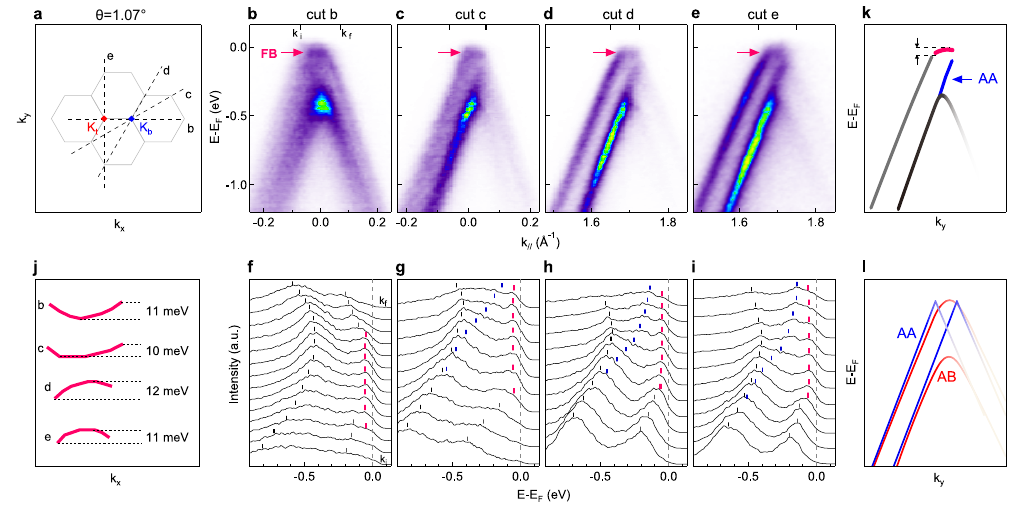}
		\caption{\textbf{The flat band observed on tBLG at twist angle of 1.07$^{\circ}$.} \textbf{a}, Schematic of the mSBZs with the dashed lines indicating the cutting directions for data shown in \textbf{b-e}. \textbf{b-e}, Dispersion images measured along  different directions to reveal the flat band (pointed by red arrows). \textbf{f-i}, EDCs from \textbf{b-e} for extracting the dispersion of the flat band. The momentum range for the EDCs is marked by tick marks in \textbf{b-e}. \textbf{j}, Extracted bandwidth of the flat band along different cutting directions. \textbf{k}, Extracted dispersions for data shown in \textbf{e}. \textbf{l}, Simulated dispersions from AA and AB stackings without considering the interlayer coupling.}
	\end{figure*}

	\begin{figure*}[htbp]
		\centering
		\includegraphics[width=16.8 cm]{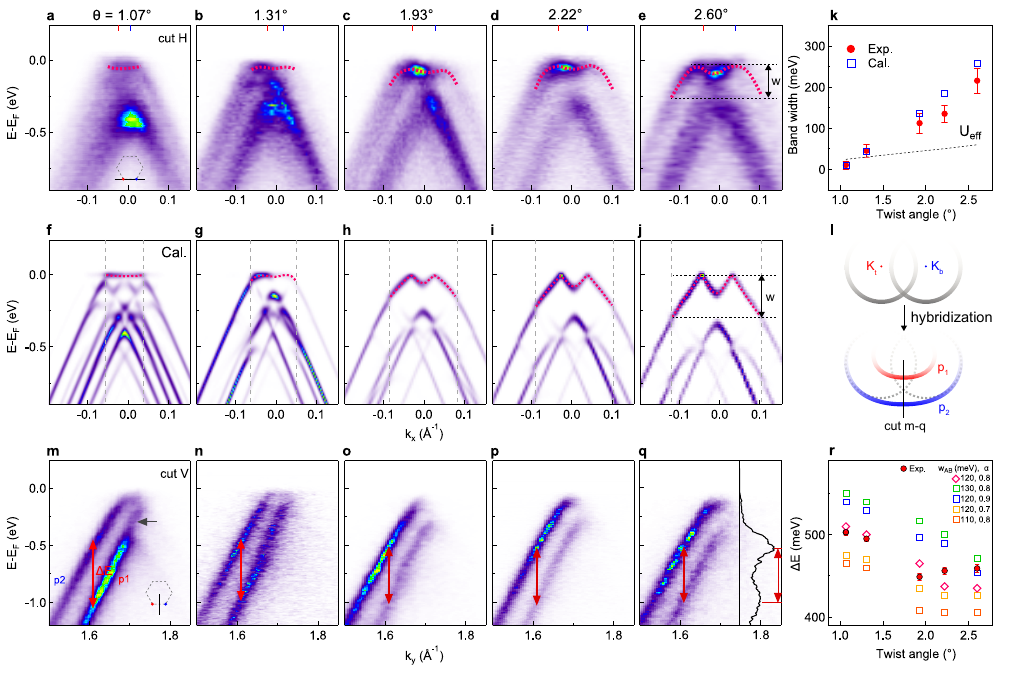}
		\caption{\textbf{Evolution of the flat band and remote bands with twist angle.} \textbf{a-e}, Dispersion images measured along the cut H direction as indicated by the black line in the inset of \textbf{a}  for twist angle from 1.07$^{\circ}$ to 2.60$^{\circ}$. \textbf{f-j}, Calculated dispersion images corresponding to \textbf{a-e}. The gray dashed lines mark where a gap is observed, and the bandwidth is extracted from dispersions within these two lines (Extended Data Fig.~4). \textbf{k}, Extracted bandwidth of the flat band from experimental results (red filled symbols) and calculated results (blue hollow squares). \textbf{l}, Schematic illustration for the interlayer interference between top and bottom graphene Dirac cones, and the resulting splitting of the remote bands into $p_1$ (red curve) and $p_2$ (blue curve). \textbf{m-q}, Dispersion images measured along the cut V direction as indicated by the black line in the inset of \textbf{m} for twist angle from 1.07$^{\circ}$ to 2.60$^{\circ}$. The inset in \textbf{q} shows EDC to extract the energy separation $\Delta E$ between $p_1$ and $p_2$. The red arrows in \textbf{m-q} indicate the size of energy separation at different twist angle. \textbf{r}, A comparison between the extracted $\Delta E$ at $ k_y = 1.61 $ \AA$^{-1}$ from \textbf{m-q} with those from theoretical calculations using different parameters of $\omega_{AB}$ and $\alpha$. Data in \textbf{k, r} are presented as fitting values from 5 samples with different twist angles, with error bars representing the standard error.}
	\end{figure*}
			
	\begin{figure*}[htbp]
		\centering
		\includegraphics[width=16.8 cm]{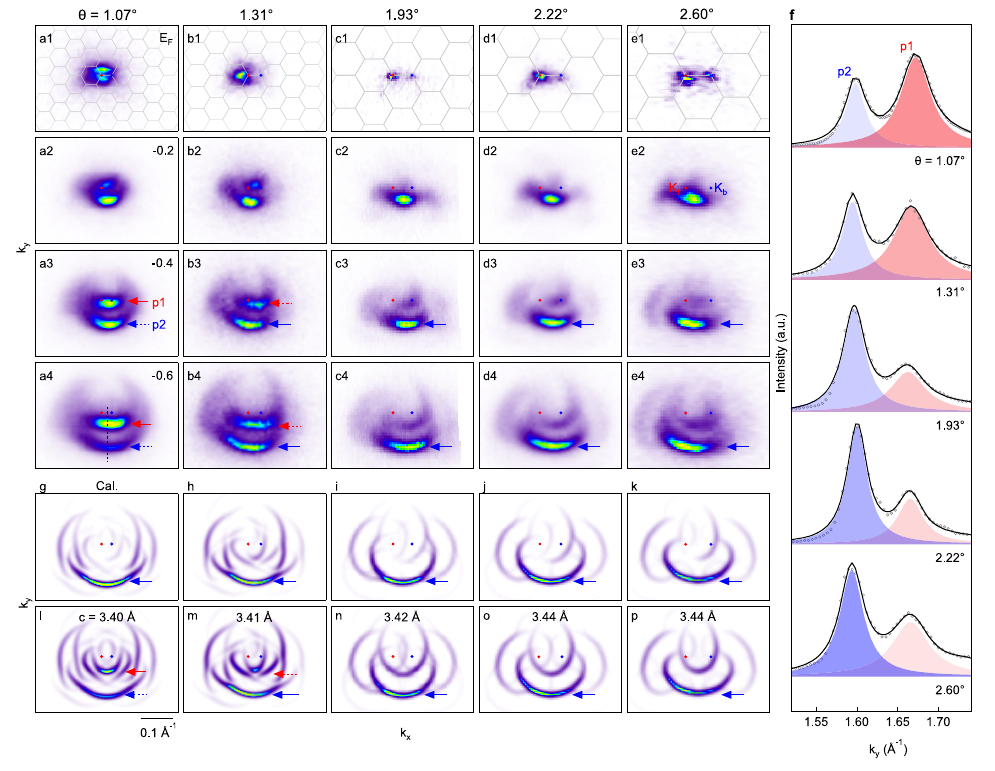}
		\caption{\textbf{Observation of spectral weight transfer from NanoARPES intensity maps.} \textbf{a1-e4}, ARPES intensity maps measured on samples with twist angle from 1.07$^{\circ}$ to 2.60$^{\circ}$. Rows from \textbf{1} to \textbf{4} correspond to energies of $E_F$, -0.2 eV, -0.4 eV and -0.6 eV respectively. \textbf{f}, Extracted MDCs at -0.6 eV as indicated by the black dashed line in \textbf{a4}. \textbf{g-k}, Calculated intensity maps at -0.6 eV with twist angle from 1.07$^{\circ}$ to 2.60$^{\circ}$ by using a fixed interlayer spacing. \textbf{l-p}, Calculated intensity maps at -0.6 eV by using different  $c$ values from 3.40 \AA ~to 3.44 \AA. The red arrows point at the pocket $p_1$ and the blue arrows point at the pocket $p_2$ to show the spectral weight transfer. }
	\end{figure*}
		
	\begin{figure*}[htbp]
		\centering
		\includegraphics[width=16.8 cm]{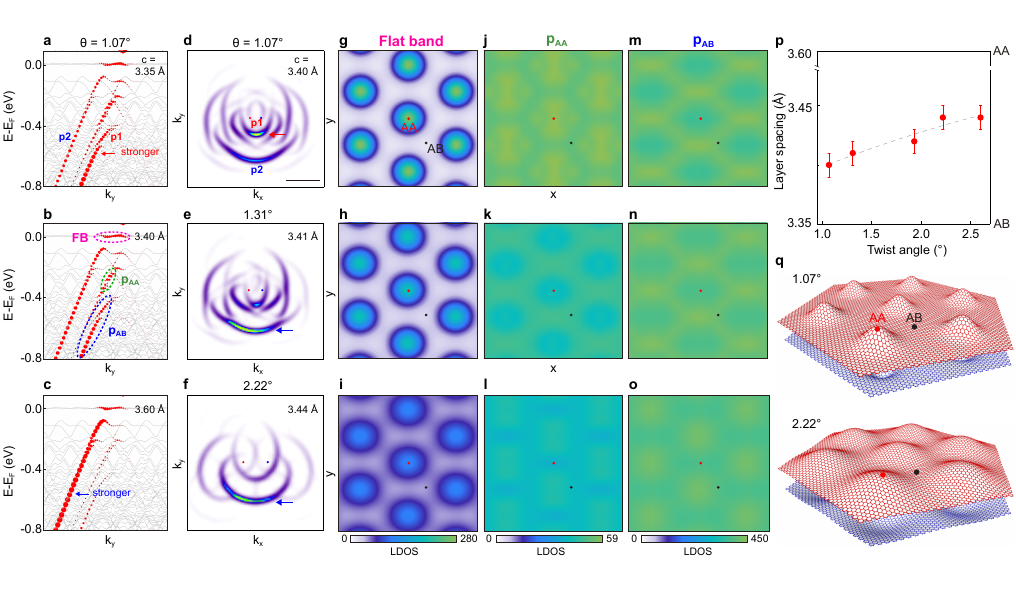}
		\caption{\textbf{Lattice relaxations revealed by theoretical calculations.} \textbf{a-c}, Calculated dispersion for twist angle of 1.07$^\circ$ by varying the  interlayer spacing $c$ = 3.35 \AA, 3.40 \AA~ and 3.60 \AA, respectively. The red, green and blue dotted ovals in \textbf{b} mark the pockets for the LDOS shown in \textbf{g-o}. \textbf{d-f}, Simulated ARPES intensity maps at -0.6 eV for twist angles of 1.07$^\circ$, 1.31$^\circ$ and 2.22$^\circ$ with  interlayer spacing of $c$ = 3.40 \AA, 3.41 \AA~ and 3.44 \AA, respectively. The scale bar is 0.1 \AA$^{-1}$. \textbf{g-o}, Calculated LDOS for the flat band (\textbf{g-i}), $p_{AA}$ pocket (\textbf{j-l}), $p_{AB}$ pocket (\textbf{m-o}) for twist angle of 1.07$^\circ$ (\textbf{g, j, m}), 1.31$^\circ$ (\textbf{h, k, n}) and 2.22$^\circ$ (\textbf{i, l, o}), respectively. \textbf{p}, Extracted interlayer spacing $c$ according to the calculation results from 1.07$^\circ$ to 2.60$^\circ$. \textbf{q}, Schematic illustration for lattice relaxations near the magic angle, where AA stacking areas are minimized. Data in \textbf{p} are presented as mean values from 3 simulation results for each twist angle, with error bars representing the standard error.}
	\end{figure*}

	\begin{addendum}
		\item[Data availability] All relevant data of this study are available within the paper and its Supplementary Information files. Source data are provided with this paper.
	\end{addendum}

\end{document}